\documentclass[useAMS,usenatbib]{mn2e}
\usepackage{color,url,graphicx, amsmath}
\usepackage{epstopdf}
\title[Jets from Quiescent Black Holes]{Constraints on Relativistic Jets in Quiescent Black Hole X-ray Binaries from Broadband Spectral Modeling}
\author[Plotkin et al.]{%
	Richard~M.~Plotkin,$^{1}$\thanks{E-mail: rplotkin@umich.edu}
         Elena Gallo,$^{1}$
	Sera Markoff,$^2$
	Jeroen Homan,$^3$  
	Peter G. Jonker,$^{4,5,6}$
	\newauthor
	James C.~A.~Miller-Jones,$^7$
	David~M.~Russell,$^{8}$
	and Samia Drappeau$^{9}$
	\\
$^{1}$Department of Astronomy, University of Michigan, 1085 South University Avenue, Ann Arbor, MI 48109\\
$^{2}$Anton Pannekoek Institute for Astronomy, University of Amsterdam, Science Park 904, 1098 XH, Amsterdam, the Netherlands\\
$^3$Kavli Institute for Astrophysics and Space Research, Massachusetts Institute of Technology, 70 Vassar Street, Cambridge, MA 02139, USA\\
$^4$SRON, Netherlands Institute for Space Research, Sorbonnelaan 2, 3584-CA, Utrecht, The Netherlands\\
$^5$Department of Astrophysics/IMAPP, Radboud University Nijmegen, Heyendaalseweg 135, 6525-AJ, Nijmegen, The Netherlands\\
$^6$Harvard-Smithsonian Center for Astrophysics, 60 Garden Street, Cambridge, MA 02138, USA\\
$^7$International Centre for Radio Astronomy Research, Curtin University, G.P.O. Box U1987, Perth, WA 6845, Australia\\
$^8$New York University Abu Dhabi, PO Box 129188, Abu Dhabi, UAE\\
$^{9}$Institut de Recherche en Astrophysique et Plan\'etologie (IRAP), 9 avenue du Colonel Roche, BP 44346-31028, Toulouse, France \\
}

\newcommand{\pasj}{PASJ}
\newcommand{\aj}{AJ}
\newcommand{\mnras}{MNRAS}
\newcommand{\apjs}{ApJS}
\newcommand{\apj}{ApJ}
\newcommand{\apjl}{ApJL}
\newcommand{\araa}{AR\&A}

\newcommand{\aap}{A\&A}

\newcommand{\nat}{Nat}   %Nature

\newcommand{\ssr}{Space Science Reviews}

\def\lesssim{\mathrel{\hbox{\rlap{\hbox{\lower3pt\hbox{$\sim$}}}\hbox{\raise2pt\hbox{$<$}}}}}
\def\gtrsim{\mathrel{\hbox{\rlap{\hbox{\lower3pt\hbox{$\sim$}}}\hbox{\raise2pt\hbox{$>$}}}}}

\newcommand{\fullsrc}{XTE~J1118+480}
\newcommand{\src}{J1118}
\newcommand{\srcasix}{A0620-00}
\newcommand{\lx}{L_X}
\newcommand{\lledd}{L_X/L_{\rm Edd}}
\newcommand{\ledd}{L_{\rm Edd}}
\newcommand{\mbh}{M_{\rm BH}}
\newcommand{\msun}{{\rm M_{\sun}}}
\newcommand{\ergs}{{\rm erg~s}^{-1}}
\newcommand{\flux}{{\rm erg~s}^{-1}~{\rm cm^{-2}}}
\newcommand{\rg}{r_{\rm g}}
\newcommand{\smbh}{SMBH}
\newcommand{\xrb}{BHXB}
\newcommand{\nh}{N_{\rm H}}

\begin{document}

%\date{Accepted 1988 December 15. Received 1988 December 14; in original form 1988 October 11}
\date{}

\pagerange{\pageref{firstpage}--\pageref{lastpage}} \pubyear{2014}

\maketitle

\label{firstpage}

\begin{abstract}
The nature of black hole jets at the lowest detectable luminosities remains an open question, largely due to a dearth of observational constraints.  Here, we present a  new, nearly-simultaneous broadband spectrum of  the black hole X-ray binary (\xrb) \fullsrc\  at an extremely low Eddington ratio ($\lx \sim 10^{-8.5}~\ledd$). Our new spectral energy distribution (SED) includes the radio, near-infrared, optical, ultraviolet, and X-ray wavebands.  \fullsrc\ is now the second  \xrb\ at such a low Eddington ratio with  a well-sampled SED, thereby providing new constraints on highly sub-Eddington accretion flows and jets, and opening the door to begin comparison studies between systems.  We apply a multi-zone jet model to the new broadband SED, and we compare our results to previous fits to the same source using the same model at 4--5 decades higher luminosity.  We find that after a \xrb\ transitions to the so-called quiescent spectral state, the jet base becomes more compact (by up to an order of magnitude) and slightly cooler (by at least a factor of two).   Our preferred model fit indicates that jet particle acceleration is much weaker after the transition into quiescence.  That is, accelerated non-thermal particles no longer reach high enough Lorentz factors to contribute significant amounts of synchrotron X-ray emission.  Instead,  the X-ray waveband is dominated by synchrotron self-Compton emission from a population of mildly relativistic electrons with a quasi-thermal velocity distribution that are associated with the jet base.  The corresponding (thermal) synchrotron component  from the jet base emits primarily in the infrared through ultraviolet wavebands.  Our results on \fullsrc\ are consistent with broadband modeling for \srcasix\ (the only other comparably low Eddington ratio  \xrb\ with a well-sampled SED) and for Sgr~A* (the quiescent supermassive black hole at the Galactic center).  The above could therefore represent a canonical baseline geometry for accreting black holes in  quiescence.  We conclude with suggestions for future studies to further investigate the above scenario.

\end{abstract}


\section{Introduction}
\label{sec:intro}
Both transient black hole X-ray binaries (\xrb s) and supermassive black holes (\smbh s) spend the majority of their time accreting at very low  rates relative to their Eddington luminosities $\ledd$,\footnote{The Eddington luminosity is the limit above which radiation pressure halts the accretion of material onto the black hole, corresponding to $\ledd = 1.26 \times 10^{38} \left[M/M_\odot\right]~\ergs$ for ionized hydrogen in a spherical geometry, where $M$ is the black hole mass.} %%
living in the so-called quiescent regime.  For \xrb s, we define quiescence  as an X-ray luminosity  $L_X  \lesssim 10^{-5}~\ledd$, corresponding to $\lesssim10^{34}~\ergs$ for an $8.5~\msun$ black hole \citep[see][]{plotkin13}.   There is general agreement that quiescent black holes accrete predominantly from some form of a radiatively inefficient accretion flow (RIAF), with X-rays emitted by a population of hot electrons.   However, there are still several open questions regarding the nature of accretion flows at such low Eddington ratios.   For instance, there is  significant debate on whether the hot electrons are primarily thermal or non-thermal, and if they are mostly inflowing or outflowing \citep[e.g.,][]{mcclintock03}.  Largely limiting our understanding is that it is unknown if quiescent black holes always launch steady collimated jets.  Therefore, current accretion models are poorly constrained regarding the degree to which jets are important, both in terms of particle acceleration and the bulk flow of the jet plasma.   

Given the low flux levels of quiescent black holes, an inherent challenge is that even the best multiwavelength datasets generally have relatively low signal-to-noise.   A natural starting point therefore is  to extrapolate trends observed at slightly higher accretion rate for ``hard state'' \xrb s ($\sim$$10^{-5} \lesssim \lx\ \lesssim10^{-2}~\ledd$; see, e.g., \citealt{remillard06, belloni10} for reviews on \xrb\ spectral states), for which higher-quality data  exists for a larger number of sources.  The dominant X-ray emission mechanism in the hard state is still  under debate.  For example, for the inflowing component, there can be a contribution from  a RIAF \citep[e.g.,][]{remillard06} and/or an efficient thin disk \citep[e.g.,][]{miller06, wilkinson09, reis10, reynolds13}, and the relative balance between the two types of flows might not be universal for every source.  Regardless, it is well-established that  hard state \xrb s are associated with compact  radio emission, which is interpreted as  optically thick synchrotron radiation from the partially self-absorbed flat spectral component of a compact relativistic jet \citep{blandford79, hjellming88, fender01}.  The compact jet typically remains unresolved in the radio, except for in a handful of cases with high (very long baseline interferometric) spatial resolution imaging \citep[e.g., GRS~1915+105; Cyg~X-1;][]{dhawan00, stirling01}.  The compact jet becomes optically thin around near-infrared (NIR) frequencies \citep[$\sim$$10^{12}$-$10^{14}$~Hz;][]{corbel02, russell13}, and synchrotron radiation from this optically thin  component can sometimes extend into the X-ray waveband \citep[e.g.,][]{markoff01a, markoff03, russell10, russell13}.  Besides emitting high-energy radiation, the  jet might also carry away the bulk of the accretion power via mechanical energy  \citep[e.g.,][and references therein]{fender03, gallo05a}.   In short, the apparent trend is that the jet becomes increasingly important in the hard state as Eddington ratio decreases.  

A major outstanding question is if the increasing importance of the jet extends all the way into quiescence.  For example, it is well-known that quiescent \xrb s have softer X-ray spectra than hard state systems \citep[e.g.,][]{kong02, tomsick03, corbel06, corbel08, plotkin13, reynolds14}.  However,  multiple  accretion scenarios can explain the X-ray spectral softening comparably well given the available (low signal-to-noise) data.  It is thus not understood if the softer X-ray  spectra  actually signify a switch in accretion properties, or if the accretion flow and jet simply evolve toward a `baseline'  as a \xrb\ approaches the quiescent state \citep[see][for details]{plotkin13}.   Further hindering our understanding is the challenge of routinely obtaining multiwavelength detections for quiescent \xrb s, largely due to the very low flux levels of these systems.  It is therefore not clear if relativistic jets always persist deep into quiescence in the first place.  For example, only two quiescent \xrb s have reliable radio detections: V404~Cyg ($\lx \sim 10^{-6}~\ledd$;  \citealt{hjellming00, gallo05}) and \srcasix\ ($\lx \sim 10^{-8.5}~\ledd$; \citealt{gallo06}).   \citet{miller-jones11} performed a deep radio survey that included 11 \xrb s in quiescence, none of which was detected in the radio.  They demonstrated that if other quiescent \xrb s launch jets with  powers and radiative efficiencies as expected from extrapolating the hard state trends into quiescence, then we can expect to  detect jet radio emission only from a select number of very nearby systems, even with our most sensitive radio telescopes.

Given these challenges, some of our best insight into quiescent black holes so far has come from multiwavelength studies of Sgr A*, the compact radio source associated with the $\sim$$4\times10^6~\msun$ SMBH at the Galactic center ($L_{X} \sim 10^{-11}~\ledd$).   \citet{falcke00} applied a relativistic jet model to the broadband spectral energy distribution (SED) of Sgr~A*, to investigate if its flat radio spectrum could signify the presence of a compact self-absorbed synchrotron jet \citep{blandford79}.  They concluded that if the radio emission is optically thick synchrotron from a compact jet, then the fraction of particles in the jet that are accelerated into a non-thermal power-law tail must be very small.   The primary  constraint leading to this conclusion is that the observed infrared (IR) spectrum implies an underlying lepton spectrum that is too steep (power law index $p>3.8$) to result from standard particle acceleration scenarios ($p=2.0-2.4$; e.g., \citealt{drury83}). The underlying particles are thus predominantly in a quasi-thermal distribution, with only a small fraction of non-thermal particles present.   However, Sgr~A* undergoes approximately daily X-ray flares that typically last for $\sim$1 hour.   During these flares, the X-ray spectrum hardens and a non-thermal radiation component emitted from within the inner few gravitational radii dominates over the quiescent X-ray emission.  Broadband spectral modeling favors scenarios where the non-thermal X-ray radiation during the flares is synchrotron emission from non-thermal leptons, likely due to sporadic particle acceleration events \citep{markoff01, dodds-eden09, dibi14}.  

The above results for Sgr A* suggest a picture where quiescence is associated with only weak, and possibly sporadic, particle acceleration in the jets.   The emission properties  of Sgr~A* when it is flaring appear to be analogous to those of hard state \xrb s \citep{markoff05a}, perhaps indicating that such a scenario might also be applicable to \xrb s as well.    However, our knowledge on the emission mechanism(s) from  \xrb s that are analogous to Sgr A* when it is not flaring is currently rather limited.  Almost all of our observational constraints on very low luminosity  \xrb s ($\lx \sim$10$^{-8.5}~\ledd$) are derived from \srcasix, because it is the only one with a well sampled SED from the radio through the X-ray wavebands.\footnote{We note that the SED of V404~Cyg  has also been well sampled in quiescence \citep{hynes09}, but its quiescent X-ray luminosity is approximately two orders of magnitude higher than \srcasix.} %%
   Interestingly, broadband modeling of \srcasix\ in quiescence indeed supports the idea of quiescent emission properties similar to Sgr~A* \citep{gallo07}.  However, we cannot determine from a single source if such a trend applies to all \xrb s.  Furthermore, \srcasix\ has been in quiescence for over 30 years (and multiwavelength coverage of its previous outburst naturally pales in comparison  to today's standards).  So pending a future outburst, it is impossible to directly compare its quiescent and hard state properties in detail.  
 
There is a strong need for additional well-sampled SEDs  of  quiescent \xrb s.  To this aim,  we recently obtained new coordinated \textit{Chandra} X-ray, \textit{SWIFT} ultraviolet (UV),  William Herschel Telescope optical and NIR, and Karl G.\ Jansky Very Large Array (VLA) radio observations of the \xrb\ \fullsrc\ (hereafter \src).   Given its nearby and well-constrained distance of $1.72\pm0.10$~kpc \citep{gelino06} and high Galactic latitude ($b=+62^\circ$; meaning that the amount of line of sight absorption is small), \src\ is one of the few known \xrb s for which it is possible to simultaneously detect both radio and X-ray emission in quiescence.   Indeed, our new VLA observation yielded the lowest-luminosity radio detection of a \xrb\ jet to date \citep{gallo14}.   With an Eddington ratio of $\lledd \sim 10^{-8.5}$, \src\ is one of our best probes of black hole accretion flows at the lowest detectable luminosities.   A special aspect of \src\ is that it has also been well-studied at higher luminosities during  previous outbursts \citep[e.g.,][]{esin01, markoff01a, mcclintock01, hynes00, hynes03, chaty03, malzac04, yuan05a, zurita06, maitra09, brocksopp10, vila12, zhang13}. 

In this paper, we apply a multi-zone jet model to our newly-assembled broadband SED of \src\ in quiescence.   The same model employed here has also been applied to \src\ in the hard state \citep{maitra09}, and also to \srcasix\ in quiescence (10$^{-8.5}~\ledd$; \citealt{gallo07}), providing a unique opportunity to \textit{uniformly} compare potential changes in accretion and jet properties as a function of Eddington ratio  within an individual source as well as to another quiescent \xrb.   Here, we focus  on the spectral modeling of these data, and we discuss these data in the context of radio/X-ray luminosity correlations in a companion paper \citep{gallo14}.   In Section \ref{sec:obs} we describe our observations and data reduction, where we add nearly simultaneous NIR, optical, and UV observations to the VLA radio and \textit{Chandra }X-ray data points.  A summary of the  jet model is included in Section \ref{sec:model}.  Results from our best model-fit are presented in Section \ref{sec:res}, which are then discussed in Section \ref{sec:disc}.   Throughout, we adopt the following parameters for \src: black hole mass $\mbh=7.5~\msun$, orbital inclination $i=68^\circ$ \citep{khargharia13}, and distance $d=1.72\pm0.10$~kpc \citep{gelino06}.  The orbital period is $P_{\rm orb}=4.08$($\pm 5\times10^{-6}$)~h \citep{torres04}.  We adopt a Galactic extinction of $A_V=0.065$~mag toward \src\ \citep{gelino06}, and a \citet{cardelli89} reddening law in the NIR through UV.   For X-ray absorption, we use $\nh=1.2\times10^{20}$~cm$^{-2}$ \citep{mcclintock03}.    The high Galactic latitude of \src\ means that its SED is virtually unabsorbed \citep[see, e.g.,][]{mcclintock03}, and our model results are not sensitive to the exact values adopted for $A_V$ and $\nh$.  All error bars on X-ray measurements and best-fit parameters are quoted at the 90\% confidence level, unless stated otherwise.  Uncertainties on flux densities at other wavebands are quoted at the 1$\sigma$ level.

\section{Observations}
\label{sec:obs}
Here, we describe the  nearly simultaneous radio, NIR, optical, UV, and X-ray observations that comprise our new SED of \src\ in quiescence.  Details on the \textit{Chandra} X-ray and VLA radio observations, which were taken through a joint \textit{Chandra}/NRAO program during \textit{Chandra} Cycle-14 (PI Gallo, Proposal 14400368), are described by \citet{gallo14}.  We only briefly summarize those observations and data here, and we describe our observations at the other wavebands in more detail.  We also include non-simultaneous IR data from Spitzer and the Wide-field Infrared Survey Explorer (WISE; \citealt{wright10}) in our SED to improve the spectral coverage (see Section~\ref{sec:irdata}).  The observations and measured flux densities  in each waveband (before applying any extinction correction) are summarized in Table~\ref{tab:sed}.

\subsection{Summary of Radio and X-ray Observations} 
We observed \src\ in the radio with the VLA in the C configuration (angular spatial resolution of $\sim$4$\arcsec$) in two overlapping 1024-MHz base bands centered at frequencies of 4.8 and 5.8~GHz.  The observations were split over two days, 2013 June 27 and 28, yielding a total of 11.3 h on source.   The data were reduced following standard procedures with the Common Astronomy Software Application \citep[CASA;][]{mcmullin07} v4.1.0.  Data from each day  were reduced and imaged separately, and then combined to create a single deep image.  A 3.2$\sigma$ peak was detected in the combined radio image, coincident with the expected position of \src.  To improve the $S/N$, we added 2.4h of integration time on source from an archival observation from 2010 November.  \src\ has a radio flux density of $4.79\pm1.45~\mu$Jy~beam$^{-1}$, which corresponds to a radio luminosity of $\nu L_{\nu} = 9.83\times10^{25}~\ergs$ at 5.3~GHz (assuming a flat radio spectrum).

The \textit{Chandra} X-ray observation was taken on 2013 June 27 (obsID 14630), with a net exposure time of 58 ks .  The target was placed at the aim point of the S3 chip on the Advanced CCD Imaging Spectrometer \citep[ACIS;][]{garmire03}.   The data were reduced following standard procedures with the Chandra Interactive Analysis of Observations ({\tt CIAO}) software, v4.5 \citep{fruscione06}.  We obtain a total of 146  counts within a circular source aperture centered on the X-ray source position (with radius = 3$\arcsec$), with an expected 15 of those being background counts (as estimated from a circular annulus with inner and outer radii of 10 and 30$\arcsec$, respectively).   The net count rate is $(2.3\pm0.3)\times10^{-3}$ counts s$^{-1}$.    Assuming a power-law with photon index\footnote{The X-ray photon index $\Gamma$ is defined as $N(E) = N_0(E/E_0)^{-\Gamma}$, where $N(E)$ is the number of photons at a given energy $E$, $N_0$ is the photon number normalization, and $E_0=1$~keV is the reference energy.}  %%
 $\Gamma=2$ (see Section~\ref{sec:res}), the absorbed 0.5-7~keV flux is $1.46(\pm 0.22)\times10^{-14}~\flux$.   In order to perform  the broadband spectral fitting (which is done in X-ray detector space; see Section~\ref{sec:res}), we extract an X-ray spectrum with the {\tt CIAO} tool {\tt specextract}.  We also create a response matrix file (rmf) and auxiliary response file (arf), applying an energy-dependent point source aperture correction to the arf  to account for the 3$\arcsec$ source aperture.   
 
 %% Observing Log
\begin{table*}
\begin{minipage}{0.9\linewidth}
\caption{Observing Log and SED}
\label{tab:sed}
%\scriptsize
%\begin{tabular}{l  cc r@{.}l c cc r@{.}l c }
\begin{tabular}{l l l l l  l l}
\hline
		Date       & %1
		Start Time\footnote{UTC is listed only for the nearly simultaneous observations.  The VLA observations started at UTC 21:30 on June 27 and at UTC 21:26 on June 28, and they lasted 7.5 h on each day.}  &  %1.5
		Telescope & %2
		Filter & %3
	 	Frequency &  %4
		Flux Dens.\footnote{Flux densities are reported prior to applying corrections for interstellar extinction.  For the Chandra observation, we report the absorbed flux  from 0.3-7~keV in erg~s$^{-1}$~cm$^{-2}$, and the quoted uncertainty is at the 90\% confidence level (all other error bars are  $\pm$1$\sigma$).}& %5
		$A_\lambda$\footnote{The extinction in each filter is calculated assuming $A_V=0.065$~mag \citep{gelino06} and a \citet{cardelli89} reddening law with $R_V=3.1$.  For the Swift/UVOT filters, we estimate the extinction using the $A_\lambda/A_V$ ratios tabulated in \citet{kataoka08}.  For the X-ray, we assume an equivalent Hydrogen absorption column density of $\nh=1.2\times10^{20}$~cm$^{-2}$ \citep{mcclintock03}. No extinction correction is applied to the non-simultaneous IR data.}\\ %6
	                        & %1
	           (UTC)             & %1.5
	                        & %2
	                        & %3
	           (Hz)     & %4
	           ($\mu$Jy)  & %5
	           (mag)   \\ %6
		\hline
		
	          \multicolumn{6}{l}{Nearly Simultaneous Observations} \\
	          \hline
		2013 June 27-28  &  21:30   & VLA   &  C-band  & $5.3\times10^9$       &  $4.79\pm1.45$ &  \\

		2013 June 27     & 20:56        &   WHT/LIRIS   & K$_s$  & $1.39\times10^{14}$      & $111.10 \pm 4.58$   & 0.008 \\
		2013 June 27     & 22:15       &   WHT/LIRIS   & K$_s$  & $1.39\times10^{14}$      & $117.31 \pm 4.98$   & 0.008  \\
		2013 June 28      & 20:53      &   WHT/LIRIS   & K$_s$  & $1.39\times10^{14}$      & $134.44 \pm 5.26$   & 0.008 \\
		2013 June 28      & 22:11        &   WHT/LIRIS   & K$_s$  & $1.39\times10^{14}$      & $128.04 \pm 5.33$   & 0.008 \\
		                                                                                          
		2013 June 27     &  21:26        &   WHT/LIRIS   & H  & $1.80\times10^{14}$          & $123.68 \pm 4.33$   & 0.012 \\
		2013 June 27     &  22:30        &   WHT/LIRIS   & H  & $1.80\times10^{14}$          & $133.63 \pm 4.97$   & 0.012  \\
		2013 June 28     &  21:26        &   WHT/LIRIS   & H  & $1.80\times10^{14}$          & $151.74 \pm 5.01$   & 0.012 \\
		2013 June 28     &  22:25         & WHT/LIRIS   & H  & $1.80\times10^{14}$          & $166.53 \pm 5.70$   & 0.012 \\
                                                                                                          
		2013 June 27     &  21:33         &   WHT/LIRIS    & J  & $2.43\times10^{14}$         & $128.50 \pm 3.63$   & 0.019 \\
		2013 June 27     & 22:37          &   WHT/LIRIS    & J  & $2.43\times10^{14}$         & $125.86 \pm 4.14$   & 0.019 \\
		2013 June 28     & 21:33          &   WHT/LIRIS    & J  & $2.43\times10^{14}$         & $125.40 \pm 3.48$   & 0.019 \\
		2013 June 28     & 22:32          &   WHT/LIRIS    & J  & $2.43\times10^{14}$         & $148.97 \pm 4.29$   & 0.019 \\
                                                                                                          
		2013 June 27      & 21:45  &   WHT/ACAM   & i$^\prime$  & $4.01\times10^{14}$   &  $87.83 \pm 0.89$ &   0.044 \\
		2013 June 28      &21:44   &   WHT/ACAM   & i$^\prime$  & $4.01\times10^{14}$   &  $80.81 \pm 0.67$ &   0.044 \\
                                                                                                          
		2013 June 27      & 21:54   &   WHT/ACAM    & r$^\prime$  & $4.86\times10^{14}$ &  $78.13 \pm 0.57$ &   0.057 \\
		2013 June 28      & 21:51   &   WHT/ACAM    & r$^\prime$  & $4.86\times10^{14}$ &  $66.59 \pm 0.48$ &   0.057 \\
                                                                                                          
		2013 June 27     & 22:04    &   WHT/ACAM    & g$^\prime$  & $6.40\times10^{14}$ &  $39.15 \pm 0.31$ &   0.079 \\
		2013 June 28     & 22:01    &   WHT/ACAM    & g$^\prime$  & $6.40\times10^{14}$ &  $32.75 \pm 0.37$ &   0.079 \\

		2013 June 27     & 19:49    & SWIFT/UVOT  & uvw1  &  $1.15\times10^{15}$  &  $3.99\pm0.99$  & 0.140 \\
		2013 June 27     & 19:39   & SWIFT/UVOT  & uvm2  &  $1.34\times10^{15}$  &  $2.24\pm0.78$  & 0.152 \\
		2013 June 27     & 19:30    & SWIFT/UVOT  & uvw2  &  $1.56\times10^{15}$  &   $1.15\pm0.61$ & 0.173 \\
		
		2013 June 27     &  16:11   &   Chandra  & ACIS  &     0.3-7 keV                       &    $ (1.46\pm0.22)\times 10^{-14}$                      &     $1.2\times10^{20}$~cm$^{-2}$ \\         
		\hline
		\multicolumn{6}{l}{Non-Simultaneous Observations} \\
		\hline
		2005 May 13       &   & \textit{Spitzer}/MIPS   & 24.0$\mu$m  & $1.25\times10^{13}$  & $<$50.0 &  \\
		2010                      &    & WISE                            & 22$\mu$m/W4                &  $1.35\times10^{13}$    &  $<$1685.6 \\
		2010                     &    & WISE                            & 12$\mu$m/W3                & $2.68\times10^{13}$  &  $<$214.9  & \\
		2004 Nov 21       &   & \textit{Spitzer}/IRAC  & 8.0$\mu$m  &  $3.75\times10^{13}$ & $59.0\pm5.9$ & \\
	  	2010			   &   & WISE                            & 4.6$\mu$m/W2                & $6.45\times10^{13}$  & $78.7\pm12.2$ & \\	
		2004 Nov 21        &   & \textit{Spitzer}/IRAC  & 4.5$\mu$m  & $6.66\times10^{13}$  & $69.0\pm6.9$ & \\
		2010                       &   & WISE                           & 3.4$\mu$m/W1                & $8.86\times10^{13}$  & $85.0\pm6.5$ & \\
\hline
\end{tabular}
\end{minipage}
%\medskip
\end{table*}

\subsection{Observations at Other Wavebands}
\subsubsection{Near-infrared and Optical}
We obtained NIR and optical observations of the counterpart to \src\ using the 4.2~m William Herschel Telescope (WHT) on La Palma (Spain). We employed two instruments,  the Long-slit Intermediate Resolution Infrared Spectrograph (LIRIS) and the auxiliary port camera (ACAM),  both in their imaging mode. Observations were obtained on 2013 June 27 and 28, where we obtained images in the K$_s$, H, J, Sloan i$^\prime$,  r$^\prime$, and g$^\prime$ filters on both nights.

For the NIR observations taken with LIRIS, we applied a 9-point dither pattern where we took two exposures of 30~s each at all of the 9 positions in the K$_s$ band, one exposure of 20~s at each position in the H-band, and one exposure of 30~s at each position in the J-band. Routines from the LIRIS data reduction pipeline {\sc Theli} \citep{schirmer13} were used to correct for the sky background and flatfield. Using information from the Two Micron All Sky Survey \citep[2MASS;][]{skrutskie06} on the position of sources detected in the individual frames, these separate frames were averaged such that we obtained two separate images of \src\ in each filter per night. We obtained a photometric calibration by using several unsaturated stars in the LIRIS field of view that are detected in the 2MASS catalog.

For the optical ACAM observations we acquired three exposures in each filter with exposure times of 240~s, 120~s, and 120~s for the g$^\prime$, r$^\prime$, and i$^\prime$ filters, respectively.  We applied standard data reduction techniques using {\sc iraf} to correct for the bias and flatfield. We combined the three images per filter to reduce the statistical error of each measurement. For the photometric calibration we used g$^\prime$, r$^\prime$, and i$^\prime$ magnitudes of several stars in the field as reported in the Sloan Digital Sky Survey \citep[SDSS;][]{york00}. 

 For each image, we list the flux densities at the effective wavelength of each filter in Table~\ref{tab:sed} (one image per night in g$^\prime$, r$^\prime$, and i$^\prime$; two images per night in J, H, and K$_s$).  The differences in the flux densities within each filter are consistent with the expected degree of periodic variability  due to orbital modulations of the secondary star.   To incorporate this systematic into the broadband spectral fitting, we use the average flux density for each of the six filters over both nights (after correcting for Galactic extinction), and then we add systematic error bars to each of the six data points at the $\pm$15\% level (the amplitude of the orbital modulations are typically $\pm$0.15-0.20~mag; \citealt{gelino06}).

\subsubsection{Ultraviolet}
We observed \src\ on 2013 June 27 with the \textit{Ultraviolet/Optical Telescope} \citep[\textit{UVOT;}][]{roming05} onboard \textit{SWIFT} \citep{gehrels04}, using the uvw1 (1554 s), uvm2 (1428 s), and uvw2 (1554 s) filters (PI Homan). Individual frames were combined using the tool {\tt uvotimsum}. In the combined images, a source was detected at the expected target position at the 2.9, 4.1, and 1.9$\sigma$ levels in the uvw1, uvm2, and uvw2 filters, respectively.    We  consider \src\ to be detected in the uvw2 filter (even though it is only at the 1.9$\sigma$ level) because it is coincident with the expected target position, and the source can be seen when visually inspecting the images.  Using the tool {\tt uvotsource}, we obtain flux density measurements of $3.99\pm0.99$ (uvw1), $2.24\pm0.78$ (uvm2), and $1.15\pm0.61~\mu$Jy (uvw2) at each filter's effective wavelength (2600, 2246, and 1928~\AA, respectively).    The systematic errors in the uvw1, uvm2, and uvw2 filters are $\pm$0.12, 0.01, 0.02~$\mu$Jy, respectively.  We correct each flux density for Galactic extinction, using the $A_\lambda/A_V$ ratios tabulated in \citet{kataoka08}.

\subsubsection{Non-simultaneous Infrared Data}
\label{sec:irdata}
\src\ appears in the all-sky data release of  WISE, which surveyed the entire infrared (IR) sky in four filters in 2010.   \src\ was detected in the W1 (3.4 $\mu$m) and W2 (4.6 $\mu$m) filters, with flux densities of $85\pm6$ ($S/N$=18.6) and $79\pm12$ $\mu$Jy ($S/N$=9.2), respectively.    There was no detection in the W3 (12 $\mu$m) and W4 (22 $\mu$m) filters, for which we adopt the 95\% confidence flux upper limits listed in the WISE catalog.  
We also include archival infrared (IR) observations from the \textit{Spitzer Space Telescope}  (PI: M. Muno; program 3289). \src\ was observed on  2004 Nov 21 with the Infrared Array Camera \citep[IRAC;][]{fazio04} and detected in both the 4.5 and 8.0 $\mu$m bands.  \src\  was also observed, but not detected, in the 24 $\mu$m band with the Multiband Imaging Photometer for Spitzer \citep[MIPS;][]{rieke04} on 2005 May 13 \citep{muno06}.  We adopt flux densities of 69 (4.5$ \mu$m) and 59 $\mu$Jy (8.0 $\mu$m),  and an upper limit of $<$50 $\mu$Jy at 24 $\mu$m, as measured by \citet{gallo07} who analyzed the same data originally presented in \citet{muno06}.   \citet{gallo07} estimated that the statistical uncertainties in the flux densities are at the 10\% level.

Quiescent \xrb s are known to experience a low-level of flux variability \citep[e.g.,][and references therein]{khargharia13, shahbaz13, bernardini14}.    \citet{muno06} estimate that a level of flux variability in the IR of $\sim$30\% could be reasonable.  From the overlap between the \textit{Spitzer} 4.5 $\mu$m band and the WISE W2 filter observations, we find a $\sim$15\% difference in flux between the two IR epochs.  We thus conservatively add systematic error bars to all IR data points at the 30\% level, in addition to the statistical uncertainties quoted above.

\section{Multi-zone Jet Model}
\label{sec:model}
The jet model employed here builds upon earlier foundations for multi-zone jets \citep[e.g.,][]{blandford79, falcke95}, and it was developed over a series of papers \citep[e.g.,][]{falcke00, markoff01, markoff03, markoff05, markoff08, maitra09}.  The earliest motivation was to study Sgr A* \citep{falcke00}, with what was a simplified version of the current model.  Now, the current model has been applied to several hard state \xrb s and low-luminosity active galactic nuclei \citep[see, e.g.,][]{markoff01, markoff05, migliari07, markoff08, maitra09}.  Below, we describe essential features of the jet model that are required to understand our current study, and a full description (and history) of the model can be found in \citet[][and references therein]{markoff05}.  Throughout, we refer to the model as the MNW05 model,  and we adopt the following notation: $\gamma_j$ refers to the bulk Lorentz factor of the jet plasma; $\beta_e$ refers to the speeds of radiating electrons, normalized to the speed of light; the corresponding electron Lorentz factors are $\gamma_e=\left(1-\beta_e^2\right)^{-0.5}$, and their energies are $\gamma_e m_e c^2$ (where $m_e$ is the electron rest mass).  All size-scales are normalized to the gravitational radius of the black hole ($\rg = G \mbh / c^2$), unless stated otherwise.  We generally use $z$ to refer to the distance of each jet zone from the black hole (the $z$-axis points along the axis of the jet), and  $r$ refers to the radius of each jet zone.

The MNW05 model is for a steady state jet, and it assumes that  the radiation is entirely leptonic\footnote{Throughout the text, we will  assume that the leptons are electrons for convenience.}  %%
 and that protons dominate the kinetic energy.  The model assumes a  maximally dominated jet, which means that the bulk internal energy (dominated by the magnetic field) is comparable to the bulk kinetic energy (dominated by protons; \citealt{falcke95}).   The total jet power is assumed to scale as $\dot{M}c^2$ at the inner edge of the accretion flow, where $\dot{M}$ is the mass accretion rate.  Within each zone of the jet, we calculate the expected flux from synchrotron radiation and SSC, which is then compared to the observed SED of \src\ in quiescence.

 The most important free parameter in the jet model is the jet power, $N_j$, which determines the initial power (normalized to $\ledd$) that is injected into the electrons and the magnetic field at base of the jet.  The very base of the jet has a cylindrical geometry (aligned along the jet axis), with a radius $r_0$ and height $z=h_0$;  we refer to this cylinder as the ``nozzle.''  The size of the nozzle is controlled by the free parameter  $r_0$, and  we fix the ratio $h_0/r_0=1.5$.    The radiating particles in the nozzle are assumed to have a quasi-thermal, mildly relativistic (Maxwell-J{\"u}ttner) velocity distribution with temperature $T_e$ ($T_e$ in the nozzle is a free parameter).  We require $T_e > \left(m_e c^2\right)/k_B=5.94\times10^9$~K because we do not consider cyclotron processes.    The plasma in the nozzle follows a gas law with an adiabatic index $\Gamma=4/3$ and has a proper sound speed $\gamma_s \beta_s \sim 0.4$ (i.e., $\beta_s^2 = \left[\Gamma-1\right]/\left[\Gamma+1\right]$). The ratio of energy injected into the nozzle  that is initially split between the magnetic field and radiating electrons is controlled by the equipartition factor $k=U_B/U_e$ ($k$ is a free parameter).  $U_B=B^2/8\pi$ is the magnetic energy density (where $B$ is the magnetic field), and $U_e$ is the electron energy density (calculated by integrating over the the entire electron distribution).

At the top of the nozzle (i.e., $z=h_0$), the jet base is allowed to freely expand laterally, which results in a longitudinal pressure gradient that accelerates the bulk plasma.  The bulk flow velocity profile is solved for exactly by the relativistic Euler equation.  Note that the conditions in the nozzle (set largely by the free parameters $N_j$, $r_0$, and $k$), combined with the above adiabatic expansion, sets all macroscopic conditions along the \textit{entire} jet (including the bulk flow velocity, electron temperature, magnetic field, equipartition factor $k$, and density profiles; see, e.g., Equation 2 of \citealt{falcke00} for analytic forms of some of these profiles).      The bulk flow acceleration is weak, typically saturating to $\gamma_j \gtrsim 2-3$ in the outer jet.

At some distance from the black hole, $z_{\rm acc}$, we assume that a significant fraction (60\%) of particles  in the jet base are accelerated into a non-thermal power-law tail.  The acceleration mechanism is unknown, but we assume that it is related to diffusive shock processes \citep[e.g.,][]{jokipii87}.  The location of the acceleration region is closely related to the location of the jet break frequency, $\nu_b$, in the SED (i.e., the frequency where synchrotron emission turns from optically thick to optically thin).  Any optically thin synchrotron (and associated SSC) extending into the X-ray waveband is primarily emitted from this acceleration zone.  As one moves into jet zones farther from the black hole, the synchrotron radiation peaks toward lower frequencies, and integrating over the outer zones gives rise to the jet's signature flat/inverted radio spectrum.  We (arbitrarily) integrate to $z_{\rm max}=3.2\times10^{12}$~cm ($2.9 \times 10^6~\rg$)  to save computation time, since jet zones at larger distances contribute radiation predominantly at frequencies below  our VLA radio data point at 5.3GHz.

 Since the details of the particle acceleration are unknown,  we take a heuristic approach to modeling the  non-thermal tail of electrons at $z>z_{\rm acc}$.  We simply assume that the non-thermal electrons follow a power-law distribution  with index p (i.e., $N\left(\gamma_e\right) \propto \gamma_e^{-p}$).  To maintain this power law against cooling losses, we assume a constant rate of particle acceleration, $t_{\rm acc}^{-1}$.  We parameterize  the microphysics of particle acceleration with the free parameter $\epsilon_{\rm sc}=\beta_{\rm sh}^2/\xi$, where $\beta_{\rm sh}$ is the relative shock velocity (in the shock frame) and $\xi$ is the ratio of the scattering mean free path to the gyroradius \citep[see, e.g.,][]{markoff08, maitra09}.  The free parameter $\epsilon_{\rm sc}$ is proportional to the particle acceleration rate \citep[see][]{markoff01a}:

\begin{equation}
\label{eq:tacc}
t_{\rm acc}^{-1} = \frac{3}{4} \epsilon_{\rm sc} \frac{eB}{\gamma_e m_e c }
\end{equation}
where $e$ is the electron charge and $B$ is the magnetic field strength at the location of the acceleration zone (all variables are in cgs units).

The minimum  particle energy of the non-thermal power law tail is set to the peak  of the Maxwell-J{\"u}ttner distribution,  according to $\gamma_{e,{\rm min}}=2.23 kT_{e,{\rm acc}}/\left(m_e c^2\right)$, where $T_{e, \rm acc}$ is the electron temperature at $z_{\rm acc}$.   The maximum particle energy, set by $\gamma_{e,{\rm max}}$, is determined by the electron energy where the particle acceleration rate ($t_{\rm acc}^{-1}$) is balanced by cooling losses.  Three sources of cooling losses are considered: (adiabatic) cooling from particles escaping the jet zone ($t_{\rm esc}^{-1}$), synchrotron cooling ($t_{\rm syn}^{-1}$), and cooling from inverse Comptonization ($t_{\rm com}^{-1}$):
\begin{equation}
\label{eq:tesc}
t_{\rm esc}^{-1}=\beta_e c/z,
\end{equation}

\begin{equation}
\label{eq:tsyn}
t_{\rm syn}^{-1}=\frac{4}{3} \sigma_{\rm T} \gamma_e \beta_e^2 \frac{U_B}{m_e c},
\end{equation}
where $\sigma_{\rm T}$ is the Thomson cross section and $U_B$ is the magnetic energy density at $z_{\rm acc}$, and

\begin{equation}
\label{eq:tcom}
t_{\rm com}^{-1} = t_{\rm syn}^{-1} \frac{U_{\rm rad}}{U_B},
\end{equation}
where $U_{\rm rad}$ is the energy density of the incident radiation providing the seed photons for inverse Comptonization.  Thus, $\gamma_{e,\rm max}$ is found by solving $t_{\rm acc}^{-1} = t_{\rm esc}^{-1} + t_{\rm syn}^{-1} + t_{\rm com}^{-1}$.    The cutoff frequency of non-thermal synchrotron radiation in the broadband spectrum is related to the values of $\gamma_{e,\rm max}$ and $B$ in the acceleration zone through $\nu_{\rm cut}=0.29 \nu_{\rm crit}$, where $\nu_{crit}=3/\left(4\pi\right) \gamma_{e,\rm max}^2 \left(e B\right)/\left(m_e c\right)$ is the critical synchrotron frequency.  The MNW05 model assumes that the cooling rate is dominated by adiabatic losses.   This assumption is important to keep in mind when interpreting our best-fits to the SED of \src.

We   refer to the combination of all jet zones closer to the black hole than the  acceleration region ($z<z_{\rm acc}$) as the ``jet base'', and we refer to regions of the outer jet at  $z>z_{\rm acc}$ as  ``post-accelerated'' zones.    The ``nozzle'' refers only  to the cylindrical component (at $z<h_0$) that is not freely expanding.  We stress that the particle distributions within the post-accelerated zones  contain a combination of both thermal and non-thermal particles, while the jet base only contains a thermal component.  In each zone, the electron temperature ($T_e[z]$) and the minimum/maximum Lorentz factors ($\gamma_{e,\rm min/max}[z]$) describing the power-law tail (in the post-acceleration regions) are adjusted to lower values as one moves away from the black hole, following the prescription set by the adiabatic expansion of the bulk plasma flow (we again note that the bulk  flow is controlled only by the conditions in the nozzle and the adiabatic expansion beginning at $z=h_0$).

As input to the model, we also  include the following  properties of the \xrb\ system: black hole mass ($\mbh$), distance ($d$),   orbital inclination ($i$), and the equivalent  Hydrogen line of sight absorbing column ($\nh$),  which we fix to the parameters listed at the end of Section~\ref{sec:intro}.   The companion star to \src\ is known to have a late-type spectral class \citep{khargharia13}.   So we include a blackbody component at a fixed temperature $T_{*}=3400$~K, and we normalize this component by assuming an emitting sphere with radius $R_*=0.56~R_\odot$.  This blackbody component contributes to $\sim$90\% of the total observed flux in each NIR filter, and $\sim$25, 40, and 70\% of the total flux in the  Sloan g$^\prime$, r$^\prime$, and i$^\prime$ filters, respectively, consistent with the expected contribution of the secondary from \citet{gelino06}.  

Finally, we do not include any thermal emission from a standard geometrically thin accretion disk, which in turn implies that there is no source of seed photons for external inverse Compton scattering.  Thus, all modeled inverse Comptonisation processes are synchrotron self-Compton (SSC).\footnote{SSC is calculated in every zone.  In practice though, the SSC emission turns out to come predominantly from zones toward the bottom of the jet base, since the photon field and electron densities are highest at those locations.} %%
   Given the observed quiescent X-ray flux of \src,   any thin disk must be relatively cool with an inner disk temperature $kT_{\rm in}\lesssim 85$~eV (generously assuming that the total observed X-ray flux accounts for only 1\% of the bolometric disk luminosity and a maximally spinning prograde black hole, and including a color correction term; see \citealt{kubota98}).  Thus, even if  a thin disk  can persist close to the innermost stable circular orbit  in quiescence, we do not have sufficient data  to constrain its properties: the blackbody emission would likely peak between the SWIFT UV and \textit{Chandra} X-ray data points, and we do not observe a sufficient number of X-ray photons to detect a soft thermal excess or reflection signatures like an Iron K$\alpha$ line.   We similarly cannot accurately constrain any  contribution of (optical) thermal radiation from the outer regions of the accretion disk, primarily due to the relative brightness of the companion star and the sampling of our SED.  We discuss these limitations in Section~\ref{sec:disc}.

\subsection{Notes on fixed and free jet model parameters}
The MNW05 model includes several input parameters, which are summarized in the notes to Table~\ref{tab:fit}.  We also note in Table~\ref{tab:fit} which parameters are fixed during the model fitting, the majority of which are constrained from observations of \src\ (e.g., $M_{\rm BH}$, $d$, $i$, $N_{\rm H}$, etc.).  
There are degeneracies among certain sets of remaining parameters, however.  These degeneracies can be difficult to disentangle when performing the model fits, especially at the low flux levels observed for \src\ in quiescence.  However, from our experience fitting this jet model to other accreting black hole systems, we have found that certain parameters tend to converge toward similar values regardless of the system being modeled (at least in the hard state), including, e.g., Cyg X-1 and GX 339--4 \citep{markoff05, maitra09}, GRO~J1655--40 \citep{migliari07}, and even  supermassive black holes like M81$^\star$ \citep{markoff08}.

For \src\ in quiescence, we therefore fix the ratio of the nozzle height to radius ($h_0/r_0$) to 1.5, and the fraction of thermal particles accelerated into the non-thermal tail in the acceleration region to 0.6.   These values are similar to values found for other systems, and most importantly are consistent with values found and/or adopted by the \citet{maitra09} fits of the MNW05 model to \src\ in the hard state (easing our goal of comparing \src\ in quiescence and the hard state).  If we were to adopt other values for these two parameters, then the remaining free model parameters would compensate to yield a fit of comparable statistical quality.  However, the broad, qualitative features of the fit would remain identical (and quantitatively, the best-fit parameters are generally fairly similar within the uncertainties on each parameter).   Therefore, as summarized in Table~\ref{tab:fit}, the main jet model parameters we explore here are the jet power ($N_j$), the equipartition at the base of the jet ($k$), the radius of the nozzle ($r_0$), the electron temperature in the nozzle ($T_e$), the particle acceleration rate (parameterized by $\epsilon_{sc}$), the location of the particle acceleration zone ($z_{\rm acc}$), and the accelerated particle power law index ($p$).

%% Best-fit Parameters
\begin{table}
\centering
\begin{minipage}{80mm}
\caption{Best-fit Model Parameters}
\label{tab:fit}
%\scriptsize
%\begin{tabular}{l  cc r@{.}l c cc r@{.}l c }
\begin{tabular}{l l l}
\hline
		Parameter (unit)      & %1
		SSC & %2
		Synchrotron  \\ %3
		  & 
		 Dominated Fit & 
		 Dominated Fit \\

		\hline
		 
		 $N_j$ (10$^{-5}$$\ledd$)   &  $1.71^{+1.52}_{-0.83}$      &   $2.67^{+4.55}_{-1.86}$ \\
  		 $k$  		                      &  $0.044^{+0.023}_{-0.013}$ &   $2.066^{+15.829}_{-1.879}$ \\
		 $r_0$ ($\rg$) 	                       & $2.3^{+1.2}_{-0.6}$               &    $2.7 \pm 0.9$  \\
		 $T_e$ ($10^{10}$K)            &   $1.735^{+0.823}_{-0.418}$  &   $0.611^{+0.286}_{-0.017}$ \\
		 $\epsilon_{\rm sc}$\footnote{Limit for SSC fit is at 95\% confidence.  $\epsilon_{\rm sc}$ is held fixed for the synchrotron fit.}             & $<1.3\times10^{-6}$      & $2.5\times10^{-3}$\\
		 $z_{\rm acc}$ ($\rg$)\footnote{90\% confidence interval is larger than the allowed range of $10\leq z_{\rm acc} \leq 125$.}           &  14.4				 	         & 20.2 \\
		 $p$                                          &   2.2    			        	         &  3\\
		 $\chi^2_r$/d.o.f.  	             &  0.59/28 & 0.54/29  \\

\hline
\end{tabular}

\medskip
The table includes the parameter space explored for the MNW05 fits to \src\ in quiescence, and descriptions for each parameter are below (including a summary of all input parameters to the model).  The first four parameters describe conditions in the nozzle, while the final three parameters describe the particle acceleration at $z=z_{\rm acc}$.  All parameters in the table are  free to vary during the spectral fits, unless stated otherwise below.  $N_j$: power injected into internal energy in the nozzle; $k$: equipartition factor, equal to the ratio of magnetic energy ($U_B$) to particle energy ($U_e$) in the nozzle;  $r_0$: nozzle radius;  $T_e$: particle temperature in the nozzle; $\epsilon_{\rm sc}$: proportional to the particle acceleration rate $t_{\rm acc}^{-1}$.  $\epsilon_{\rm sc}=\beta_{\rm sh}^2/\xi$, where $\beta_{\rm sh}$ is the relative shock velocity and $\xi$ is the ratio between the scattering mean free path to the gyroradius; $z_{\rm acc}$: distance of acceleration zone  from the black hole; $p$: power-law index for accelerated leptons ($N_e \sim \gamma_e^{-p}$).  $p$ is held fixed to $p=2.2$ for the `SSC' fit, and $p=3$ for the `synchrotron' fit (see Section \ref{sec:res});  $\chi^2_r/\mathit{\nu}$: reduced $\chi^2$ for $\mathit{\nu}$ degrees of freedom.  We fix the following parameters during the fit: $\nh=1.2\times10^{20}$~cm$^{-2}$; black hole mass $\mbh=7.5~\msun$; orbital inclination $i=68^{\circ}$; distance $d=1.72$~kpc; the ratio of nozzle height to radius $\left(h_0/r_0\right)=1.5$; the fraction of nozzle particles accelerated into a power-law tail = 0.6; the temperature of the companion star $T_*=3400$~K;  and the radius of the companion star $R_*=0.56~R_\odot$.  Emission from the jet is calculated out to a distance $z_{\rm max}=3.2\times10^{12}$~cm ($2.9\times10^6~\rg$) from the black hole.
\vspace{-5mm}
\end{minipage}
\end{table}

%% Physical Parameters
\begin{table}
\centering
\begin{minipage}{80mm}
\caption{Physical Parameters}
\label{tab:phys}
%\scriptsize
%\begin{tabular}{l  cc r@{.}l c cc r@{.}l c }
\begin{tabular}{l l l l l l}
\hline
		Parameter      & %1
		\multicolumn{2}{c}{SSC-dominated} & % 2-3
		\multicolumn{2}{c}{Synchrotron-dominated} \\
		
		  (unit)  & %1
		   ($z=h_0$) &  ($z=z_{\rm acc}$) & %
		  ($z=h_0$) &  ($z=z_{\rm acc}$) \\

		\hline
		\multicolumn{5}{l}{\textbf{Parameters calculated by model}} \\
        		 $B~(10^4$G) &        13.28 &         2.03 &        56.42 &         8.13 \\
		 $n_e$ (10$^{15}$ cm$^{-3}$) &      $2.2$ &      $0.075$ &      $2.2$ &     $0.063$ \\
   	     	 $\gamma_j$ &      ... &         1.72 &      ... &         1.73 \\
	       	$\gamma_{e,\rm{max}}$ &      ... &          147 &      ... &        17723 \\

		\multicolumn{5}{l}{\textbf{Model input parameters}} \\
		$k$ &        0.044 &        0.025 &        2.066 &        1.811 \\
           	$r~(r_g)$ &          2.3 &          6.5 &          2.7 &          8.0 \\
   		$T_e~(10^{10}$K) &        1.735 &        1.152 &        0.611 &        0.403 \\

\hline
\end{tabular}

\medskip
\textit{Parameters calculated by model} are reported at the top of the nozzle ($z=h_0$) and at the particle acceleration zone ($z_{\rm acc}$).  \textit{Model input parameters} are key free parameters describing the nozzle.  We repeat their best-fit values from Table~\ref{tab:fit} here, and we also report their values at $z=z_{\rm acc}$ to illustrate how these parameters change along the jet.   $B$: magnetic field; $n_e$: number density of electrons in each zone; $\gamma_j$: bulk Lorentz factor of the plasma.  We do not report a value for $\gamma_j$ at $z=h_0$ because the plasma has only just begun expanding.  $\gamma_j$ saturates to $\gamma_j \sim 3$ in the outer jet in both the SSC- and synchrotron-dominated fits.  $\gamma_{e,\rm max}$: the maximum electron Lorentz factor after particle acceleration (see \S \ref{sec:model}).  No $\gamma_{e,\rm max}$   is reported at the top of the nozzle because there is no particle acceleration within zones at $z<z_{\rm acc}$;  $k$: equipartition factor; $r$: radius of jet zone; $T_e$: electron temperature of the thermal electron component.
%\vspace{-5mm}
\end{minipage}
\end{table}

  \begin{figure*}
\includegraphics[scale=0.9]{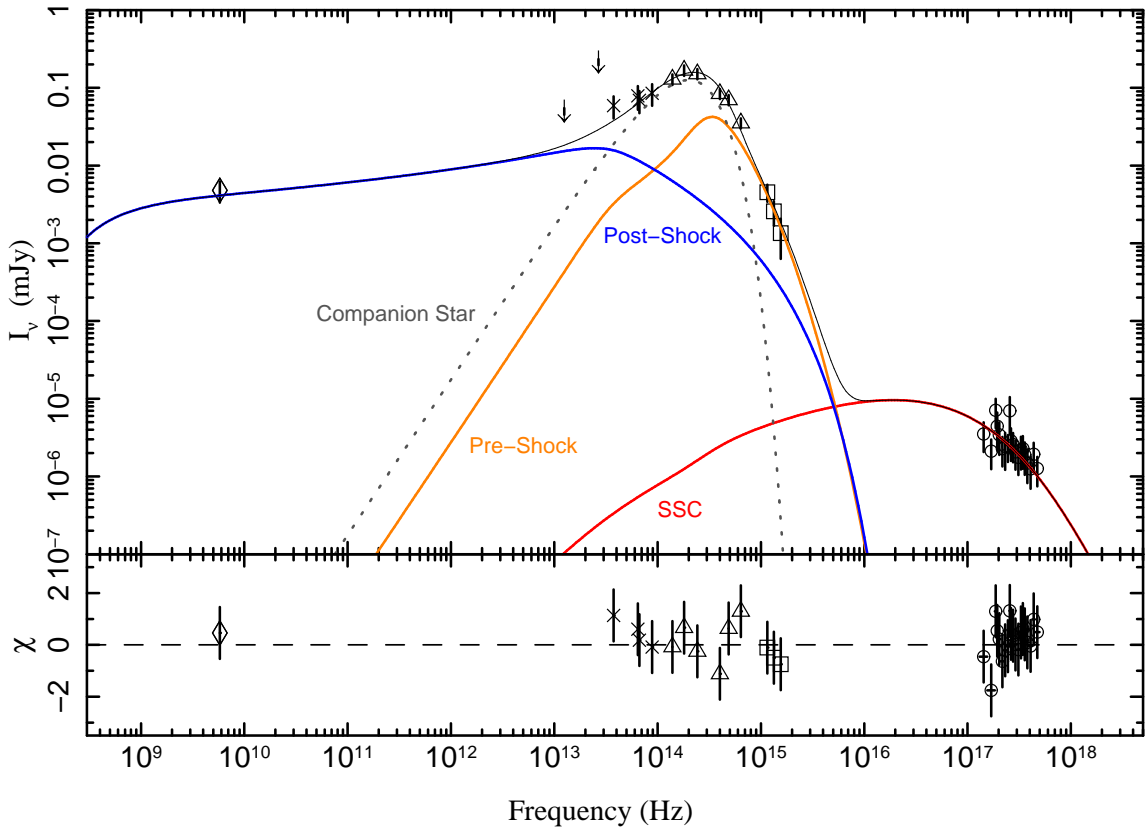}
 \caption{Broadband spectrum and best-fit model.  Symbols are for the radio (diamond), (non-simultaneous) IR (arrows for upper limits; crosses for detections), NIR and optical (triangles), UV (squares), and X-ray (circles) data points.  The jet model is fit in X-ray detector space, with the best-fit shown with the black solid line.  Also shown are the contribution from the pre-shock (thermal) synchrotron emission (orange line), the post-shock (non-thermal) synchrotron emission (blue line), SSC (red line), and the companion star (dotted black line). }
 \label{fig:sscfit}
 \end{figure*}

  \begin{figure*}
\includegraphics[scale=0.9]{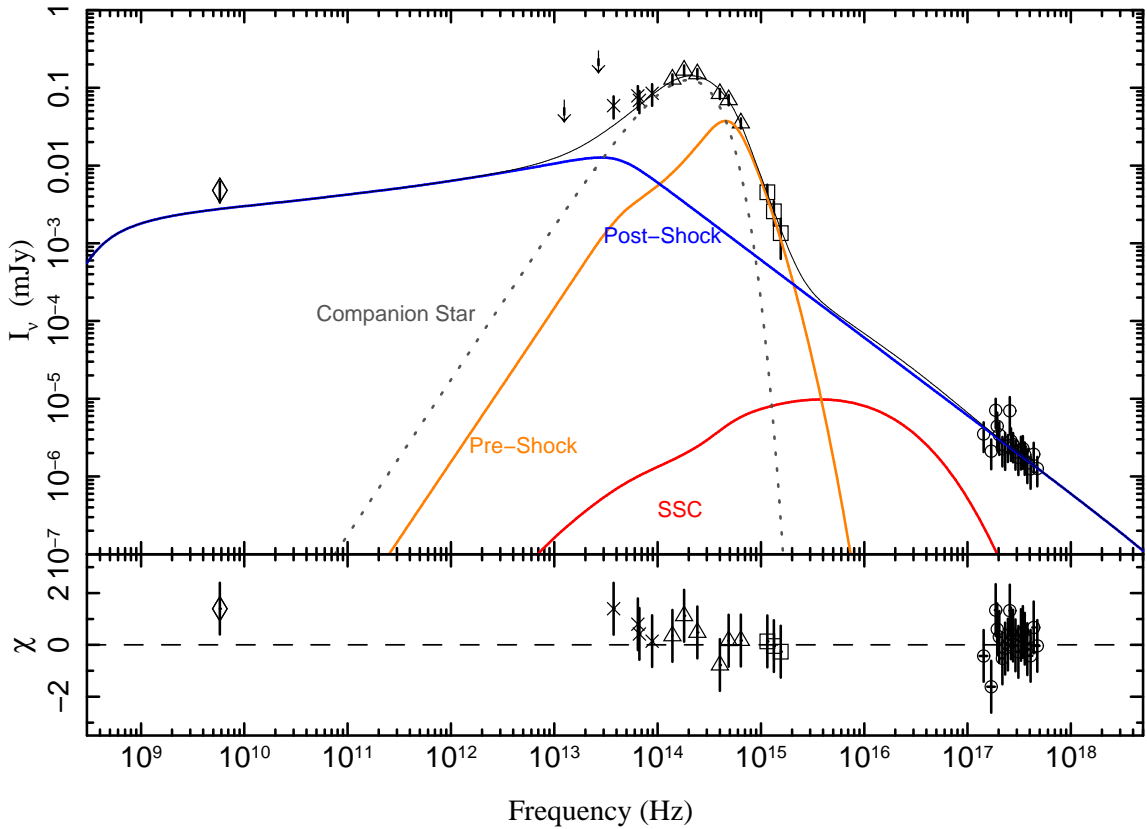}
 \caption{Best-fit model where X-rays are dominated by post-shock (non-thermal) synchrotron emission (see Section~\ref{sec:resjetfit}).  Symbols and lines have the same meaning as in Figure~\ref{fig:sscfit}.}
 \label{fig:jetfit}
 \end{figure*}

\section{Results}
\label{sec:res}
Before performing detailed modeling,  we first confirm that the X-ray spectrum is typical for a quiescent \xrb\ by fitting just the X-ray spectrum with a (phenomenological) power law modified by Galactic absorption ($\nh$ is fixed to the value in Section~\ref{sec:intro}).    The X-ray spectrum is fit  within the  Interactive Spectral Interpretation System \citep[{\tt ISIS};][]{houck00} v1.6.2-10 using Cash statistics \citep{cash79}, and we find  a best-fit photon index $\Gamma=2.02\pm0.41$.  This photon index is consistent with other quiescent \xrb s \citep{plotkin13}, and nearly identical to a 2002 \textit{Chandra} observation of \src\ in quiescence \citep[$\Gamma=2.02\pm0.16$;][]{mcclintock03}.    With this photon index and our adopted values for $\nh$ and distance,  we estimate an intrinsic (i.e., unabsorbed) 1-10~keV X-ray luminosity of $L_X=4.5\times10^{30}~\ergs$ using the Chandra Portable, Interactive Multi-Mission Simulator \citep[PIMMS;][]{mukai93}.   The implied Eddington ratio is $\left( \lledd\right) = 10^{-8.5}$.  
  
Next, we fit the  jet model to the broadband  spectrum described in Section~\ref{sec:obs}.   An important feature of the model is that the predicted spectrum is forward folded through the X-ray response, and the fit is performed within {\tt ISIS}.  Fitting in ``X-ray detector space''  allows better control over instrument-related systematics, and also a direct comparison of the goodness of fit via, e.g., $\chi^2$ statistics, at all  observed wavelengths in the broadband spectrum.   The NIR, optical, and  UV data points are corrected for extinction prior to performing the fitting (see  $A_\lambda$ in Table~\ref{tab:sed}), and the X-ray absorption is applied within {\tt ISIS} during the fit (fixing $\nh$ to our adopted value).     We  fit the model to the data by minimizing $\chi^2$ (we also try several different sets of initial guess parameters to ensure that the fit is not converging toward a local minimum).  The {\tt ISIS} script {\tt conf\_loop} is used to iteratively search for 90\% confidence intervals on each free parameter (i.e., $\Delta \chi^2=2.71$ for one  parameter of interest), and the fit is updated if a better solution is found during the confidence interval search.    The fit converges toward small values of $\epsilon_{\rm sc}$, such that the high-energy (post-acceleration) synchrotron  cutoff ($\nu_{\rm cut}$)  falls below the X-ray waveband.  With such a low value of $\nu_{\rm cut}$, it is not possible for the accelerated tail of particles to  contribute significant amounts of optically thin synchrotron to  the NIR through X-ray wavebands. We therefore cannot  constrain the slope of the optically thin synchrotron component from the data, so we instead fix $p=2.2$\footnote{The  spectral index $\alpha_{\nu }$ $\left(f_{\nu} \sim \nu^{\alpha_\nu}\right)$ for optically thin synchrotron emission from a power-law distribution of relativistic particles is related to $p$ as $\alpha_\nu = -\left(p-1\right)/2$.  Similarly, the photon index $\Gamma=\left(p+1\right)/2$.   A value of $p=2.2$ is often assumed for compact jet emission.  The other best-fit model parameters are not sensitive (within their 90\% confidence intervals) to the exact $p$ value chosen from $p=2-3$.   }  %%
and refit the model in order to more tightly constrain the other free parameters.  

The best fit to the broadband spectrum is shown in Figure~\ref{fig:sscfit}, with the best-fit parameters listed in Table~\ref{tab:fit}.  Physical parameters calculated by the code (e.g., magnetic field,  electron number density, bulk Lorentz factor, etc.) are reported in Table~\ref{tab:phys}, with values listed at the top of the nozzle ($z=h_0$) and at the acceleration zone ($z=z_{\rm acc}$).  We include values in both zones to illustrate how these physical parameters evolve along the jet.  We also repeat some key input parameters to the nozzle from Table~\ref{tab:fit} (e.g., zone radius, electron temperature, etc.), to illustrate how those input parameters evolve along the jet.   We obtain a reduced $\chi^2_{\rm r}=0.59$ for 28 degrees of freedom.  Since we (somewhat arbitrarily) assign systematic uncertainties to some data points to account for potential variability, we do not necessarily expect $\chi^2_{\rm r}$ to be close to unity.  However,  the minimum $\chi^2_{\rm r}$ still is useful for determining the best-fit \textit{relative} to the searched parameter space,  and we also require the fit residuals to not show any obvious trends as a function of frequency upon visual inspection (see bottom panel of Figure~\ref{fig:sscfit}). We  cannot strongly constrain the exact value of $\epsilon_{\rm sc}$ from the data, so we report its 95\% confidence upper limit in Table~\ref{tab:fit}, which is also the value adopted in Figure~\ref{fig:sscfit}.   We note, however, that our upper limit on $\epsilon_{\rm sc}$ is robust, since larger values would allow too much optically thin synchrotron to contribute to the UV and X-ray wavebands, resulting in a poorer statistical fit (although see below).   Since the X-ray emission is modeled predominantly  by SSC, we  refer to this as the SSC-dominated fit.

\subsection{Exploring Parameter Space: Synchrotron Dominated X-rays}
\label{sec:resjetfit}
Since we cannot directly identify the synchrotron cutoff frequency $\nu_{\rm cut}$ from the data, we further explore the $\epsilon_{\rm sc}$ parameter  here.  We note that $\epsilon_{\rm sc}$ is the only parameter we were not able to adequately investigate from the combination of different sets of  initial guess parameters and running {\tt conf\_loop} above.   We first search for a solution where  $\nu_{\rm cut}$ falls inside the X-ray band, which would result in both SSC and optically thin synchrotron contributing to the observed X-rays.   The model would not converge to such a fit.   We also could not find an acceptable fit where $\nu_{\rm cut}$ falls between the UV waveband (as in Figure~\ref{fig:sscfit}) and the soft X-ray waveband.

Next, we explore if the other extreme is possible where X-rays  are dominated by optically thin synchrotron radiation emitted by the accelerated (non-thermal) particles.   To do so, we force  $\nu_{\rm cut}$ to fall above the X-ray waveband by (arbitrarily) fixing $\epsilon_{\rm sc}=0.0025$.  In this case, we can directly constrain the spectral slope of the synchrotron emission (emitted by the post-accelerated non-thermal electrons) by using the best-fit X-ray photon index $\Gamma$, and we fix $p=3$.   Interestingly, we obtain a  fit of similar statistical quality as the above SSC dominated fit,  with a reduced $\chi^2_{\rm r}=0.54$ for 29 degrees of freedom (see Figure~\ref{fig:jetfit}, Table~\ref{tab:fit}, and Table~\ref{tab:phys}).  We note that for this `synchrotron'-dominated fit, the nozzle electron temperature $T_e$ is almost non-relativistic (i.e., it is close to  the imposed lower boundary in the nozzle, and it becomes non-relativistic at larger distance from the black hole), and the equipartition factor $k$ is poorly constrained.  The SSC- and synchrotron-dominated fits are compared in the next section.

\section{Discussion}
\label{sec:disc}

We applied a multi-zone jet model to a new broadband spectrum of \src\  in quiescence ($\lledd \sim 10^{-8.5}$), which is only the second \xrb\ at such a low Eddington ratio to have a radio detection and an SED sampled well enough to attempt broadband modeling.   The same model has previously been applied to \src\ in the hard state  during its 2000 and 2005 outbursts \citep{maitra09}, and also to \srcasix\ in quiescence \citep{gallo07}.     We obtain  two model fits for \src\ in quiescence, of comparable statistical quality,  that can explain the observed  X-rays either as  SSC (emitted by a quasi-thermal  population of  relativistic electrons) or as optically thin synchrotron emission (from an accelerated non-thermal population of electrons).     Before describing the physical differences between these two fits, we first discuss their similarities.   The origin of X-ray emission from both hard state and quiescent \xrb s is a highly debated topic.  However, by highlighting the common features among these two extremes, we can (partly) transcend this debate and obtain fairly robust insight into the nature of jets launched in quiescence.

Both fits converge toward relatively small nozzle radii $r_0\sim 2-3~\rg$.  This nozzle radius is smaller than the best-fit values for \src\ in the hard state \citep[$10-20~\rg$;][]{maitra09}, and comparable to the best-fit value for \srcasix\ in quiescence \citep[$3.9_{-0.1}^{+2.2}~\rg$;][]{gallo07}.  When compared to the hard state, both fits also converge toward lower electron temperatures (\citealt{maitra09} found $T_e \sim 4 \times 10^{10}$~K).   Thus, in the context of this jet model, as a \xrb\ transitions from the hard state into quiescence, the jet base becomes more compact (perhaps by an order of magnitude), and it evolves toward  a lower temperature (by at least a factor of two).    

It is interesting that we can explain the bulk of the IR--UV SED  with only the combination of emission from the jet and companion star, while completely neglecting any blackbody radiation from the outer regions of the accretion flow.   We do not take this result as evidence for the absence of an outer disk, as our reasons for excluding the disk component are largely systematic (see Section~\ref{sec:model}).  An outer disk  is likely present, but decomposing its emission from other radiative processes in quiescence likely requires an even better sampled SED with higher $S/N$, and/or high-resolution spectroscopy.  For example, excess UV emission over the expected  contribution from the companion star is often detected from quiescent \xrb s (see \citealt{hynes12} and references therein; although also see \citealt{hynes09} who did not find a significant UV excess in the relatively luminous quiescent SED of V404~Cyg).    Some of this UV excess is likely thermal  radiation from the outer disk, as UV spectroscopy often reveals broad emission lines  (including for both \src\ and \srcasix; \citealt{mcclintock03, froning11}).  However, a caveat is that explaining typical UV excess fluxes purely via thermal blackbody radiation requires a hotter and/or more compact emission region than expected for the outer regions of quiescent accretion flows \citep[e.g.,][]{mcclintock95, mcclintock03, froning11, hynes12}, and complex geometries may be required \citep[e.g.,][]{mcclintock03}.    Our model fits on \src\ suggest that jet-related synchrotron radiation (from a relativistic population of quasi-thermal electrons in the jet base) could also substantially contribute to the UV waveband and should be taken into account.  Note that synchrotron radiation from the jet base can also explain the bulk of the UV excess from \srcasix\ in quiescence  (\citealt{gallo07, froning11}; although see \S 4 of \citealt{froning11} for other potential scenarios).

The NIR and optical wavebands are largely dominated by the companion star in our fits to \src.  However, we find an IR excess (relative to the contribution from the companion star), which we account for as synchrotron radiation predominately from the accelerated (non-thermal) electron component.    However, since our IR data points are non-simultaneous, we cannot exclude the possibility that a circumbinary disk may instead contribute to some of the excess IR emission, especially in the Spitzer 8-$\mu$m band \citep[e.g.,][]{muno06, wang14}.   If a circumbinary disk is relevant to the IR, then in order to self-consistently also explain the jet radio emission,  the jet break ($\nu_b$) would likely need to fall at lower frequencies than in either fit (i.e., the location of the acceleration region, $z_{\rm acc}$, would be located farther from the black hole; see \citealt{gallo07}).  Unfortunately, due to lack of data between between the radio and IR wavebands, neither fit constrains $z_{\rm acc}$ to high-precision.  Thus, resolving this discrepancy would require higher frequency radio and/or sub-mm observations to better constrain the radio spectral index and to attempt to locate the jet break.  Searches for IR variability  would also be helpful (the excess IR emission should be variable if it comes from the jet).

\subsection{An SSC origin for X-ray emission in quiescence?}
\label{sec:discssc}
Although formally the $\chi^2_r$ values of  the SSC- and synchrotron-dominated fits are comparable, we argue here that the SSC-dominated fit  is more believable (in the context of the MNW05 model).   First, the uncertainties on the best fit parameters $N_j$ and  $k$ are especially large for the synchrotron-dominated fit, providing some hesitation on the fit quality.  Plus, we obtained the synchrotron-dominated fit by assuming a high particle acceleration rate, i.e.,we fixed the parameter $\epsilon_{\rm sc} \propto t_{\rm acc}^{-1}$  to a large value.  Our motivation for doing so was to help explore the full parameter space, and to investigate if quiescent black hole jets can  efficiently accelerate electrons to high $\gamma_e$.  However, if we refit the synchrotron-dominated fit and allow $\epsilon_{\rm sc}$ to vary as a free parameter, its value does not change significantly, which may suggest that the model simply converged toward a local minimum.

Perhaps more important, and regardless of our above suspicions, the synchrotron-dominated fit also appears to be approaching a parameter space that  violates some assumptions behind the MNW05 model.  For one,  the model assumes a relativistic fluid in the nozzle and jet base, yet the electron temperature in the nozzle for the synchrotron-dominated fit is uncomfortably close to the imposed limit $T_{e,\rm min}=5.94\times10^9$~K.  As the electron temperature follows the adiabatic expansion of the bulk flow, the temperature eventually drops below $T_{e,\rm min}$ (e.g., see $T_e$ at $z=z_{\rm acc}$ in Table~\ref{tab:phys}).

Furthermore, the accelerated particles must have a power law index $p\sim3$ in order to explain the X-ray data.  Such a soft particle spectrum is unlikely to be injected by a shock.  Therefore, if synchrotron radiation from non-thermal electrons indeed dominates the X-ray waveband in quiescence, then that radiation must be synchrotron cooled (i.e., the synchrotron cooling break, $\nu_{\rm cool}$, lies below the X-ray waveband in quiescence).  Below, we confirm that synchrotron cooling losses are unlikely negligible in  the X-ray waveband for the synchrotron-dominated fit.  To illustrate this point we consider  only the acceleration zone ($z_{\rm acc}$), since the synchrotron flux at high-energies from non-thermal electrons will be largest in this region where the magnetic field is strongest.   Electrons in $z_{\rm acc}$ (where   $B\sim8 \times 10^4$ G; see Table~3) that emit synchrotron in the X-ray band ($\approx$$2.4\times10^{17}$~Hz) have $\gamma_{e} \approx 1560$.   Inserting these numbers into Equations~\ref{eq:tesc}--\ref{eq:tcom}, we find that synchrotron losses account for 92\% of all cooling losses at $\gamma_e \approx 1560$ (adiabatic losses account for the remaining 8\%, and losses from inverse Comptonisation are negligible).

For the synchrotron-dominated fit, radiative losses from synchrotron cooling are therefore not  negligible in the X-ray waveband.  This means that the initial particle spectrum (at lower $\gamma_e$) must be harder than $p\sim3$.  That is, the spectrum should initially have $p\sim2$ (since the particle index should change by $\Delta p\sim 1$ above and below the synchrotron cooling break), and then the spectrum would soften to $p\sim3$ at higher electron Lorentz factors when synchrotron cooling becomes significant.  This  effect on the shape of the particle spectrum is not included in the MNW05 model, which assumes that adiabatic losses always dominate (we note, however, that some of this spectral evolution could be ``hidden'' by the optically thick portion of the jet).    The MNW05 model also assumes that the particle spectrum is similar across the entire outer jet.  That assumption was motivated by multiwavelength campaigns on jets from active galactic nuclei (where the evolution of the spectral index can be spatially resolved along the jet) that do not show spectral evolution over distances much greater than the cooling lengths \citep[e.g.,][]{jester01}.  However, since we cannot spatially resolve \xrb\ jets in quiescence, we cannot directly test this assumption via observations of \src.

{Despite the above, we can still draw some qualitative conclusions from the synchrotron-dominated fit (i.e., in the case that particle acceleration is efficient and X-rays are synchrotron cooled).  In this case, the synchrotron cooling break must fall below the X-ray waveband.  If the cooling break falls above the IR waveband (i.e., at frequencies above the jet break $\nu_b$), then the optically thin synchrotron emission would initially have a flatter spectral index between the jet break  and the cooling break.  That  implies that $\nu_b$ should fall at a lower frequency than shown in Figure~\ref{fig:jetfit}.  In turn, there is less room for the jet to account for any IR-excess, making the idea of a circumbinary disk more likely.\footnote{We note that the NIR-optical emission is still dominated by the companion star.  The synchrotron contribution to the UV emission would unlikely change much, since the UV is dominated by the thermal jet base.}  %%
However, the above is only one possibility, as it could instead be the case that the synchrotron cooling break  falls below the IR band and within the optically thick portion of the jet.  In that case, the synchrotron contribution to the IR band from the non-thermal electrons would not change by a large amount.  Unfortunately, we cannot draw strong quantitative conclusions (besides our concerns on the quality of the model fit, we also must bear in mind that our IR constraints are non-simultaneous).  However, we do note that the above issues do not affect the properties of the jet base, and they therefore do not alter our conclusions at the beginning of Section~\ref{sec:disc}.

So far, our hesitation to favor the synchrotron dominated fit is primarily due to a concern that the fit approaches a parameter space that is inconsistent with some of the underlying assumptions behind the MNW05 model.  However, the synchrotron cooled X-ray scenario becomes slightly less appealing when also considering our results on the radio/X-ray luminosity correlation in \citet{gallo14}.  With the new data point of \src\ in quiescence  (at $\lledd \sim 10^{-8.5}$), \citet{gallo14} demonstrate that \src\ exhibits a tight, non-linear radio/X-ray luminosity correlation over five decades in X-ray luminosity, of the form $L_{\rm r} \propto L_{X}^{0.72 \pm 0.09}$.   The slope of the non-linear correlation is suggestive of radiatively inefficient X-ray processes.  More specifically, to explain this slope, the X-ray luminosity should  scale approximately quadratically (depending on the radio spectral index) with the normalized mass accretion rate $\dot{m}=\dot{M}/\dot{M}_{\rm Edd}$  \citep[e.g.,][]{markoff03}.  However, synchrotron cooled emission from non-thermal electrons scales \textit{linearly} with $\dot{m}$ \citep[e.g.,][]{heinz04}, which would result in a steeper luminosity correlation with a slope almost twice as large \citep{heinz04, yuan05}.   So, if \src\ were to transition from radiatively inefficient X-rays  in the low-hard state to synchrotron cooled X-rays in quiescence, as would be implied by the synchrotron-dominated fit, then the slope of its radio/X-ray luminosity correlation should also steepen in quiescence \citep{yuan05}.  We do not observe such a steepening of the slope in the radio--X-ray luminosity plane.  Therefore, in order for the synchrotron-dominated fit to be correct,  the X-rays would have to transition to being synchrotron cooled at relatively low luminosities (so that the  steepening of the slope is not noticeable even at $\ledd \sim 10^{-8.5}$), or the expected scalings of radio and X-ray luminosity  depend on other parameters in addition to  $\dot{m}$.

We conclude that the SSC-dominated fit is likely a more believable representation of the data, in the context of the MNW05 model.  However, we  again stress that the above concerns are only relevant to the post-acceleration jet zones and \textit{not} the jet base or nozzle.  Therefore, despite the above, we still consider our qualitative results that the nozzle becomes smaller and cooler in quiescence to be robust.  To properly investigate the feasibility of efficient particle acceleration (i.e., high $t_{\rm acc}^{-1}$ and high $\gamma_{e,\rm max}$) and synchrotron cooled X-rays would require adjustments to the model out of the scope of this paper (see \citealt{yuan05} for the application of a jet model including synchrotron cooling losses to an optical/UV/X-ray SED of \src\ in quiescence).    We also stress that this issue is not a concern for the SSC-dominated fit.  For the SSC-dominated fit, particle acceleration is very weak ($\gamma_{e,\rm max}\sim147$) and the magnetic field is smaller, so that adiabatic losses always dominate, consistent with our assumptions.  Therefore, we are more confident in the physical parameters found by the SSC-dominated fit, provided that the X-rays indeed are predominantly SSC emission from a relativistic distribution of thermal electrons.  Unfortunately, it is not trivial to predict the expected radio/X-ray correlation if \xrb s switch to SSC-dominated X-rays from a thermal electron population, so it is unclear at this point whether the SSC-dominated fit is consistent with the \citet{gallo14} non-linear correlation.  We expand more on the SSC-dominated fit in the next subsection.

\subsection{Jets in Quiescence}
\label{sec:discjets}
We  begin this section by comparing similarities between our jet model and  other types of accretion flows, in order to highlight robust results.   \citet{mcclintock03} undertook an optical, UV,  and X-ray campaign on \src\ in quiescence in 2002, for which they use an advection dominated accretion flow  \citep[ADAF;][]{ichimaru77, narayan94, abramowicz95}  to explain the X-ray spectrum.  An ADAF is a type of RIAF, where the black hole is fed by a radiatively inefficient two-temperature plasma with very weak Columb coupling between ions.  The radiative cooling timescale is longer than the free fall time into the black hole, resulting in under-luminous X-rays compared to a standard  thin accretion disk, and a large fraction of accretion energy is  advected directly into the black hole.\footnote{Note that ADAFs are prone to developing various types of instabilities \citep[e.g.,][]{narayan95}, and other variants of RIAFs are also possible (one example is the convection-dominated accretion flow, CDAF;  \citealt{narayan00, quataert00}).  In particular, as illustrated by the adiabatic inflow-outflow solution \citep[ADIOS;][]{blandford99}, it is very plausible that a significant fraction of the accretion energy is instead  carried away as mechanical energy in the form of an outflow, and not advected directly into the black hole \citep[also see e.g.,][and references therein, for the role that an outflow may play]{fender03}.}   %%
 Even without including an outflow, or with the benefit of  radio constraints, \citet{mcclintock03}  conclude that the X-rays are emitted via SSC (at least for $\lesssim 100$~keV photons; see \citealt{esin97}) by a  population of hot electrons, and they speculate that a non-thermal electron component could also be relevant.  Of course, a major  difference between the ADAF model and our jet model is that the jet model explicitly attributes the X-ray emission to an outflowing quasi-thermal component.        Also, an attractive feature of   jet models is that the outflow   self-consistently explains the radio emission, since  the physical conditions at the base of the jet (responsible for the optical/UV thermal synchrotron and X-ray SSC) determine the properties in the outer regions of the outflow (responsible for the radio emission). \citet{yuan05} applied a hybrid ADAF/jet model to the SED from \citet{mcclintock03}, and they also prefer jet-dominated X-rays in quiescence (although they argue for synchrotron from non-thermal electrons).
 
According to the SSC-dominated fit,  our main conclusion is that the outer jet of \src\ experiences less efficient particle acceleration in quiescence compared to the hard state (i.e., the accelerated non-thermal electron tail does not reach  high Lorentz factors).   This conclusion is consistent with the picture described for Sgr~A* in Section~1 (which also appears to undergo weaker particle acceleration and have a non-thermal SSC contribution to the quiescent X-rays; e.g., \citealt{falcke00, markoff01}).  A similar conclusion was also reached for \srcasix\ in quiescence, for which the same jet model converged toward similar best-fit parameters as we find for \src\ \citep{gallo07}.    Furthermore, the \xrb\ Swift J1357.2-0933 was recently suggested to have the lowest  quiescent  X-ray luminosity of any known \xrb\ \citep{armaspadilla14}, making it  suitable for comparisons to \src\ and \srcasix.  Swift J1357.2-0933  has a very  steep NIR--optical spectrum ($\alpha_{\nu}=-1.4$) in quiescence, which is also consistent with  synchrotron radiation from a thermal distribution of electrons in a weak jet  \citep{shahbaz13}.  Thus, the best-fit model for \src\ in Figure~\ref{fig:sscfit} may indeed represent the baseline accretion/jet properties for quiescent black holes.  

We note that the best-fit $k$-values are fairly low for the SSC-dominated fit.   That could cause some concern, because a very small magnetic energy might violate the assumption in the MNW05 model of a maximally dominated jet.  However, the mechanism(s) in which energy redistributes itself in the jet launching zone (i.e., the nozzle) are not well understood, so it is difficult for us to quantify if the small $k$ values are unphysical or not.  It is potentially interesting that the best-fit to \srcasix\ in quiescence with the same model also prefers a small-$k$ \citep{gallo07}; the small equipartition between magnetic field and electron energy densities could therefore be hinting at an interesting phenomenological property of quiescent black holes worth focusing on in the future.

Considering the above (and in the context of the SSC-dominated fit), it could be the case  that an important difference between quiescent and hard state \xrb s is the degree to which an accelerated electron component contributes to the high-energy radiation.   As \xrb s fade into quiescence, the  jet base  becomes  less magnetically dominated, cooler, and more compact, and the maximum energy of any accelerated electrons becomes smaller.   The net result may be a weaker outflow that does not develop the necessary structures to efficiently accelerate particles \citep[see, e.g.,][]{polko10}.  The X-rays in turn switch from being a combination of optically thin synchrotron emission \citep[e.g.,][]{markoff01a, plotkin12} and/or emission associated with the hot flow \citep[e.g.,][]{esin01, yuan05a} in the hard state, to becoming dominated by SSC off the outflowing (quasi)-thermal jet in quiescence.   Such a switch could also be the cause of the observed X-ray spectral softening as a \xrb\ transitions from the hard state into quiescence \citep{plotkin13}.   If all  \xrb\ accretion flows and jets evolve toward a similar baseline in quiescence, then it may also be natural to expect diverse accretion properties in the hard state (depending on the strength of the non-thermal electron component).  For example, for \src, the slope of the NIR-optical spectrum in the \textit{hard} state has been observed to range from very steep \citep[$\alpha_{\nu}\sim -1.4$;][]{russell13} to values more typical of optically thin synchrotron radiation \citep[$\alpha_{\nu}\sim -0.8$;][]{hynes06,russell13} at different epochs, which could be reflecting different levels of particle acceleration.   In addition to the above arguments that \xrb s eventually reach a quiescent baseline, the idea for less variety in quiescence might  also be supported by multiwavelength observations that track transient \xrb s through the radio/X-ray luminosity plane as they fade into quiescence following an outburst \citep[e.g.,][]{jonker10, jonker12, ratti12},  To test the above idea further, it would be helpful to have a prediction on the expected slope of radio/X-ray luminosity correlations if X-rays are dominated by SSC (from a thermal electron distribution), and also more simultaneous radio and X-ray constraints on highly quiescent black holes (to learn if all quiescent black holes have similar radio to X-ray flux ratios).

\section{Summary}
The recent  detection of radio emission from \src\ deep in quiescence \citep{gallo14}  provides new, much needed constraints on accretion flows and their jets at the lowest detectable Eddington ratios ($\lx \sim 10^{-8.5}~\ledd$).  Currently, the only other \xrb\ with a radio detection and well-sampled SED at such a low Eddington ratio is \srcasix\ \citep{gallo06, gallo07}.   From the combination of both sources, we can start to lay a foundation to ultimately learn if all quiescent \xrb s have similar accretion properties, if relativistic jets always persist at the lowest detectable Eddington ratios, and the degree to which accelerated non-thermal electrons are energetically important.   These constraints are relevant across the entire black hole mass scale, as most SMBHs also likely accrete in the quiescent regime (if they are not completely dormant).

We undertook a coordinated multiwavelength campaign to assemble a broadband spectrum for \src\ in quiescence, including radio (VLA), NIR/optical (WHT), UV (Swift), and X-ray (Chandra) observations.  We then applied a multi-zone  jet model to the broadband spectrum \citep{markoff05} to constrain the physical parameters of the system, and to tease out the dominant emission mechanism(s) in each waveband. The same model has also been applied to \srcasix\ in quiescence \citep{gallo07} and to \src\ at higher luminosities in the hard state \citep{maitra09}, allowing us to make uniform comparisons.  

We can adequately model the entire spectrum by including radiation only from the outflowing jet, and flux from the companion star.  As in the hard state, the radio emission is attributed to the sum of multiple zones of self-absorbed synchrotron emission from the outer jet \citep[e.g.,][]{blandford79}.  As \src\ fades into quiescence, we determine that its jet base becomes more compact (by up to an order of magnitude) and slightly cooler (by at least a factor of two).  Meanwhile, in our preferred model fit, the jet base also becomes less magnetically dominated, and particle acceleration becomes less efficient (i.e., non-thermal electrons in the outer regions of the jet do not attain high Lorentz factors).   Ignoring the companion star, the optical/UV emission is  thermal synchrotron emission from a mildy relativistic population of quasi-thermal electrons in the jet base, and the X-rays are corresponding SSC.  The particle acceleration is too weak for non-thermal electrons  to  contribute significant amounts of high-energy radiation.  We do not require a circumbinary disk \citep[e.g.,][]{muno06} in the IR, or thermal emission from the outer disk in the optical/UV, but in reality both components could contribute emission at some level.

The above scenario is consistent with results on \srcasix, and also with broadband modeling of Sgr~A* (using a similar jet model; e.g., \citealt{falcke00, markoff01}).  We thus speculate that \src\ and \srcasix\ could represent a canonical baseline for quiescent black hole accretion flows and jets.  The overall structure could also be similar for hard state \xrb s.  However, at higher luminosities, there is likely an increased flux of disk seed photons near the jet base for external inverse Compton scattering, and reflection off the accretion disk will also be more important.  The  primary difference inferred from our work might also be  the degree to which hard state jets can accelerate a non-thermal tail of electrons.  With stronger acceleration, non-thermal particles may contribute more radiation to the high-energy wavebands.

In the future,   well-sampled SEDs for more quiescent \xrb s are clearly needed, which can currently be obtained only for very nearby (and ideally high Galactic latitude) systems.  In particular, additional high-resolution UV spectra (e.g., with  COS on HST) would  be helpful for constraining the disk contribution to the UV \citep[e.g.,][]{froning11}.   Observational constraints on the high-energy cutoff would also be extremely useful, for which coordinated UV and X-ray observations (of unabsorbed systems) could be a promising avenue.  Curvature between the UV and X-ray band could be indicative of a cooling break.  Hard X-ray constraints would also be useful if the high-energy cutoff falls at hard X-ray energies.    Besides the high-energy cutoff, the other poorly constrained parameter from the jet model is the location of the jet acceleration zone, $z_{\rm acc}$, which is important  for understanding the jet's energetics.  Improving constraints on $z_{\rm acc}$ requires more coverage from the sub-mm through IR, which could be achieved with ALMA for some sources, and/or the James Webb Space Telescope in the future.   We note that, while we await such observations, there is already a positive outlook to more tightly constrain quiescent jet properties (with current data) through  improvements in the theoretical modeling.  The next generation of the jet model employed here will self-consistently derive the flow solution from the jet base to acceleration zone ($z_{\rm acc}$) for a given set of initial conditions in the inner accretion flow \citep{polko10, polko13, polko14}, which will significantly reduce the number of free parameters.
  
\section*{Acknowledgments}

We thank Marianne Heida for her help with the reduction of the NIR WHT images, and we thank Neil Gehrels and the Swift team for approving and scheduling the Swift/UVOT observations.  Support for this work was provided by the National Aeronautics and Space Administration through Chandra Award Number GO3-14036X issued by the Chandra X-ray Observatory Center, which is operated by the Smithsonian Astrophysical Observatory for and on behalf of the National Aeronautics Space Administration under contract NAS8-03060.  JMJ is supported by an Australian Research Council (ARC) Future Fellowship (FT140101082) and also acknowledges support from an ARC Discovery Grant (DP120102393).  SD acknowledges funding support from the french Research National Agency: CHAOS project ANR-12-BS05-0009.  This research has made use of software provided by the Chandra X-ray Center (CXC) in the application package CIAO.  The William Herschel Telescope is operated on the island of La Palma by the Isaac Newton Group in the Spanish Observatorio del Roque de los Muchachos of the Instituto de Astrof'sica de Canarias.  The National Radio Astronomy Observatory is a facility of the National Science Foundation operated under cooperative agreement by Associated Universities, Inc.  This publication makes use of data products from the Wide-field Infrared Survey Explorer, which is a joint project of the University of California, Los Angeles, and the Jet Propulsion Laboratory/California Institute of Technology, funded by the National Aeronautics and Space Administration.  This work is based in part on archival data obtained with the Spitzer Space Telescope, which is operated by the Jet Propulsion Laboratory, California Institute of Technology under a contract with NASA.

%\bibliographystyle{mn2e}
%\bibliography{ref}

\bsp

\label{lastpage}

\end{document}